\documentclass[11pt,preprint]{aastex}

\received{}
\revised{}
\accepted{}
\ccc{}
\cpright{}{}

\shorttitle{M13 Faint C and N Abundances}
\shortauthors{Briley et al.}

\newcommand{\kms}{km~s$^{-1}$}

\newcommand{\teff}{$T_{eff}$}

\newcommand{\cabund}{[C/Fe]}
\newcommand{\nabund}{[N/Fe]}
\newcommand{\ciso}{$^{12}$C$/^{13}$C}

\begin{document}

\title{The Chemical Inhomogeneity of Faint M13 Stars: C and N Abundances
\altaffilmark{1}}

\author{Michael M. Briley\altaffilmark{2}, 
Judith G. Cohen\altaffilmark{3},
Peter B. Stetson\altaffilmark{4}}

\altaffiltext{1}{Based in large part on observations obtained at the
W.M. Keck Observatory, which is operated jointly by the California
Institute of Technology, the University of California and NASA.}

\altaffiltext{2}{Department of Physics and Astronomy, University of Wisconsin
Oshkosh, 800 Algoma Boulevard, Oshkosh, WI 54901}

\altaffiltext{3}{Palomar Observatory, MS 105-24, California Institute of Technology,
Pasadena, CA 91125}

\altaffiltext{4}{Dominion Astrophysical Observatory, 5071 West Saanich Road,
Victoria, BC V9E 2E7, Canada}

\begin{abstract}
Building upon earlier observations which demonstrate substantial star-to-star
differences in the carbon abundances of M13 subgiants,
we present new Keck LRIS spectra reaching more that 1.5 mag below the
M13 main-sequence turn-off (to V $\approx$ 20).
Our analysis reveals a distribution of C abundances similar to that found
among the subgiants,
implying little change in the compositions of the M13 stars at least through
the main-sequence turn-off.
We presume these differences to be the result of some process operating
early in the cluster history.

Additional spectra of previously studied bright M13 giants have been obtained
with the Hale 5-m.
A comparison of C abundances derived using the present methods and
those from the literature yield a mean difference of 0.03$\pm$0.14 dex for
four stars in common with \citet{1996AJ....112.1511S} and 0.14$\pm$0.07 dex for
stars also observed by \citet{1981ApJS...47....1S} (if one extreme case is removed).
We conclude that the lower surface C abundances of these luminous giants
as compared to the subgiants and main-sequence stars
are likely the result of mixing rather than a difference in our
abundance scales.

NH band strengths have also been measured for a handful of the most
luminous M13 turn-off stars.
While molecular band formation in such stars is weak, significant star-to-star
NH band strength differences are present.
Moreover, for the stars with both C and N measurements, differences
between stars in these two elements appear to be anticorrelated.

Finally, the most recent C and N abundances for main-sequence,
main-sequence turn-off, and subgiant stars in 47 Tuc, M71, M5, and the present M13 data
are compared.

\end{abstract}


\section{INTRODUCTION}

It has been known since the early 1970's that the otherwise
indistinguishable members of
any given Galactic globular cluster (GC) exhibit significant
star-to-star variations in surface abundances of certain light elements
(most notably C and N, as well as O, and often Na, Al, and
Mg)\footnote{Note we are excluding
$\omega$ Cen and M22, both of which appear to have experienced
some degree of self-enrichment.}.
However, while the abundance patterns commonly observed point to an
origin in proton capture nucleosynthesis \citep{1990SvAL...16..275D,
1995PASP..107.1177L, 1996ApJ...464L..79C}, identification of the specific
reaction site(s) and a full theoretical description of the abundance
modifying process(es) remain uncertain.
As has been pointed out in numerous reviews \citep[see][]{1987PASP...99...67S,
1994PASP..106..553K, 1998IAUS..189..193D} two possibilities exist:

First, the present day cluster stars may have modified their own surface compositions
through some mixing process not included in standard
models (i.e., an ``in situ'' scenario).
By far the most promising candidate site in this regard is the region
above the H-burning shell after first dredge-up in evolving cluster giants, where conditions
for partial CN and possibly ON-cycle reactions exist \citep[see][for
one of the earliest treatments]{1979ApJ...229..624S}.
Subsequent circulation of this material into the stellar envelope via meridional
currents or turbulent diffusion \citep[for example][]{2003ApJ...593..509D}
will result in decreasing C
abundances and increasing N with evolutionary state as has been
observed along the red giant branches (RGB) of several metal-poor
clusters \citep[see][for classic examples]{1982ApJS...49..207C,
1983ApJ...266..144T, 1990ApJ...359..307B}.
Moreover, the operation of such a mechanism can at least
qualitatively explain the O and Mg versus Na and Al anticorrelations 
found among the most luminous red giants in several clusters
\citep[e.g.][and references therein]{1998AJ....115.1500K}.

Common to all models of this process is the prohibition of
``extra mixing'' by the molecular weight gradient left behind by the inward excursion
of the convective envelope during first dredge-up.
Only after the molecular weight discontinuity has been destroyed by the
outward moving H-burning shell, an event marked by the RGB luminosity
function (LF) bump, is mixing expected to take place \citep[][and others]{1979ApJ...229..624S,
1998A&A...332..204C}.
This theoretical prediction appears to be borne out by observations of
decreased Li abundances following the LF bump in NGC 6752 \citep{2002A&A...385L..14G}
and similar drops in \ciso\ seen by \citet{2003ApJ...585L..45S} in
NGC 6528 and M4.

However, this cannot be the entire picture.
As early as 1978, it was noted by \citet{1978ApJ...223L.117H}
that the subgiant branch (SGB) and likely the main-sequence (MS)
stars of 47 Tuc also possessed star-to-star differences
in CH and CN band strengths.
This has been most recently followed in 47 Tuc to $\approx 2.5$ mag below the
MS turn-off (MSTO) by \citet{2003AJ....125..197H}.
An analysis of their observed CN and CH band strengths yields factors
of 10 variations in N anticorrelated with factors of 3 differences in C
\citep{Briley-2004}, matching those found among the more evolved members.
Such CN and CH (N and C) variations have also been shown to exist
among the MS, MSTO, or SGB stars of NGC 6752, M71, and M5
\citep*{1991ApJ...381..160S, 1999AJ....117.2434C, 2002AJ....123.2525C}
Moreover, star-to-star variations in Na, Al, O, and Mg,
similar to those found among the luminous cluster stars, have
been identified among the SGB and MSTO stars of 47 Tuc \citep{1996Natur.383..604B},
NGC 6752 \citep{2001A&A...369...87G}, M71 \citep{2002AJ....123.3277R}, and
M5 \citep{2003AJ....125..224R}.
Although the various correlations and anticorrelations among these
elements suggest the presence of material exposed to proton-capture reactions,
such stars lie well below the LF bump and, particularly in the case of
the MS stars, no mechanism is known for circulating significant quantities of
CN(O) nucleosynthesized material to their surfaces.

Thus the second possible origin of the GC abundance variations - they
have been set in place before RGB ascent and are due to the operation
of some mechanism early in the cluster history (sometimes referred to as
a ``primordial'' scenario).
A number of possibilities exist as are discussed extensively by
\citet{1998MNRAS.298..601C}, including: that the proto-cluster
gas was inhomogeneous in these elements 
(a true primordial origin), that there was an extended period of star formation
of sufficient duration to allow some low-mass stars to form with material
ejected from more massive already-evolved cluster asymptotic giant branch
(AGB) stars, or that the
present day cluster stars have accreted AGB ejecta onto their surfaces
after their formation.
The appeal of AGB stars as sites of the proton capture
nucelosynthesis lies in their ability to modify the cluster gas in light elements
\citep[including C, N, O, Al, Na, and Mg - see][]{2001ApJ...550L..65V} while not
altering the abundances of heavy elements.

As the reader has likely noted, observational evidence exists for both
mixing and early contamination scenarios, which has 
led many investigators to conclude that the compositions
of the cluster stars we observe today are not the result of one or the
other scenario exclusively, but rather both.
Unfortunately, this leads to difficulties in disentangling the contributions
of each process among the more luminous cluster stars - a problem that
can only be reconciled by exploring the compositions of a cluster's
stars to the MSTO and below.
Clearly, abundance trends found among a cluster's MS stars reflect
the original makeup of the bright giants, while deviations from this
``baseline'' composition are likely the result of mixing.
This was recently demonstrated in the case of M13 by
\citet{2002ApJ...579L..17B} (hereafter BCS02) - that a large spread in
C abundances exists among the SGB stars of M13, which presumably
reflects star-to-star variations in C abundances set early in the cluster history.
However, the SGB C abundances also appear larger than those
found by other investigators among the more luminous M13 stars,
implying the operation of a mixing mechanism on the RGB
which has reduced surface C abundances.

In the present paper, we return to M13 and extend our sample more
than two magnitudes fainter to include MS stars.
In addition, we have also obtained spectra of M13 bright giants
observed in earlier studies to verify our abundance scale.
Our results confirm those of BCS02 - that a primordial
spread in the distribution of light elements exists in M13 which has
further been modified during RGB ascent.
Measurements of the
3360\AA \ NH bands also were obtained for a handful of the
more luminous stars in our sample.
N abundances calculated from these bands suggest a C versus
N anticorrelation at the level of the MSTO.

\section{OBSERVATIONS}
\subsection{THE FAINT STAR SAMPLE}
The initial sample of stars in M13 was aimed to produce subgiants
at the base of the RGB.
It consisted of those stars from the photometric database
\citep[described by][]{1998PASP..110..533S,2000PASP..112..925S}
located more than
150 arcsec from the center of M13 (to avoid crowding) with
$16.9<V<17.35$ and with $0.86 < (V-I) < 0.96$ mag.
A slitmask with 0.7 arcsec wide slitlets, narrower than normal to
enhance the spectral resolution and minimize contributions 
from adjacent stars in these crowded fields, was designed using JGC's software
from this sample and used in May 2001 with LRIS \citep{1995PASP..107..375O} at Keck. 
For this slitmask, as for all those used for the M13 stars,
the red side of LRIS was set to include the NaD lines and H$\alpha$.  
We used the highest possible dispersion, 0.64\,\AA/pixel (29 \kms/pixel) or 1.9\,\AA/spectral
resolution element there, to facilitate
radial velocity confirmation of cluster membership.
Given that the radial velocity of M13 is --246 \kms, distinguishing field stars from
cluster members is then straightforward.

The blue side of LRIS \citep{LRIS} was used with the 600
line/mm grism blazed at 5000~\AA.
The detector for LRIS-B at that time was a 2048x2048 CCD not
optimized for UV response.
The spectra covered the range from $\sim$3400 to 5000~\AA, thus
including  the strongest CN band at 3885\,\AA\ and the G band of CH
at 4300\,\AA, with a resolution of $\sim$4~\AA\ (1.0~\AA/pixel). 
Two additional slitmasks were defined from this sample and used
in May 2002 during less than ideal weather conditions
for 6 exposures of 4800 sec each. 
The spectra were dithered by moving the stars along the length of
the slitlets by 2 arcsec  between exposures. 
These spectra are part of those presented in BCS02.

Because of the crowded fields, in addition to the intended stars
some slitlets contained additional stars bright enough to 
provide suitable spectra, and these were utilized as well.  
As might be expected from the luminosity
function, most of the secondary sample consists of stars at
or just below  the main-sequence turnoff.
Hence subtraction
of sequential exposures was not possible, and they were reduced
individually using Figaro \citep{figaro} scripts,
then the 1D spectra for each object were summed. 

Based on the serendipitous main sequence stars found in the 2002 observations
(see the plots in BCS02), 
we decided to try to reach  main sequence stars well below
the turnoff in M13, sufficiently faint to be cool enough to have detectable CH bands.
The criteria used to define the sample from the
photometric database were 19.3 $<$ I $<$ 19.7, V-I within 0.06 mag
of the main sequence of M13, taken as 1.26+0.28(I-19.4), 
and located more than 200 arcsec from the center of M13.
A single slitmask with 0.8 arcsec wide slitlets
was designed and used at Keck with LRIS June 26, 2003.
The blue spectra cover the full range from the
atmospheric cutoff to 5000~\AA, with 1.0 ~\AA/pixel and a spectral
resolution of $\sim$ 4~\AA.  Four exposures totalling 4200 sec
were obtained.
The new very high quantum efficiency detector for LRIS-B
consisting of two 2kx4k Marconi CCDs, with 15 $\mu$\ pixels and
a readout noise of 4.0 e$^-$, was completed and installed into LRIS in
June 2002,
and so was available for these observations.
The high UV throughput of LRIS-B with this new sensitive detector for the
first time enabled us to reach the NH bands in the brighter of these stars
with some precision.
The locations of the faint program stars on the M13 color-magnitude
diagram (CMD) are shown in Figure \ref{fig_1}.

\subsection{THE BRIGHT STAR SAMPLE}

There are published surveys \citep{1981ApJS...47....1S,
1996AJ....112.1511S} in which CH indices have been used to determine
\cabund\ values for large samples of the highest
luminosity giants in M13. 
However, our Keck/LRIS sample of low luminosity stars in M13 has no
overlap with these earlier works.
To ensure that the merger of our data for faint stars in M13
with these published datasets for CH band strengths in M13 giants is
valid, we need to verify the consistency of the different measurements of
the CH indices and resulting abundances.
To demonstrate this, we obtained new blue spectra of a small sample of 
bright giants with published CH band strengths from earlier
studies, and remeasured their CH indices with the same procedures used
for the lower luminosity M13 stars of our main sample (as described below).
These spectra were taken in April and May, 2003 
at the Hale Telescope on Palomar Mountain during
observing runs intended primarily for other projects. 
The  blue channel of the Double Spectrograph \citep{1982PASP...94..586O}
was used with a 1200 line/mm grating and a Loral 512x2788
15$\mu$\ pixel CCD, giving 0.55~\AA/pixel with a spectral resolution
of 1.9~\AA\ for a 1 arcsec slit.

\section{ANALYSIS}
\subsection{THE FAINT STAR SAMPLE}

Our analysis essentially repeats that of BCS02 and is fully
described in \citet{2001AJ....122..242B} (hereafter BC01)
and the reader is referred to these works for details.
To summarize: strengths of the 4350\AA ~ CH (G) bands of our program stars were
measured via the I(CH) index of
\citet{1999AJ....117.2428C,1999AJ....117.2434C} - an index
which compares the flux removed by the G-band to the adjacent continuum on both sides.
The resulting indices, calculated using bandpasses corrected for
the radial velocity of M13, are plotted for
the sample of faint stars as a function of I magnitude in Figure \ref{fig_2}.
The decrease in CH band strengths near I $\approx$ 18 is due to the
higher temperatures of the MSTO stars (as pointed out by BC01).
However, among the fainter MS stars in the sample (near I $\approx$ 19.5),
the surface temperatures have dropped by roughly 300K and
again a large and significant scatter in CH band strengths is apparent.
The one sigma error bars plotted for the present sample have been determined
entirely from Poisson statistics in the molecular-band and continuum spectral
windows.

In a similar manner, the strength of absorption by the 3350\AA \ NH
band was measured in spectra of the more luminous members of
the Keck MS/MSTO sample using the double sided logarithmic
$s_{NH}$ index as defined in \citet{1993PASP..105.1260B}.
The resulting indices (and one sigma Poisson error bars) are also
plotted in Figure \ref{fig_2}.
This marks the first time NH bands have been observed among
such faint stars in a globular cluster.
Spectra of two MSTO stars exhibiting differing NH band strengths,
and two MS stars with differing CH band strengths are shown in Figure
\ref{fig_3}.

In order to relate the observed indices to the underlying \cabund, we
employ a series of synthetic spectra based on MARCS
\citep{1975A&A....42..407G} model atmospheres.
Our models are those used in BC01 and BCS02
and based on the 16 Gyr $[Fe/H]$ = $-$1.48 O-enhanced isochrone
grid of \citet{2001ApJ...556..322B}.
The locations of the model points on the M13 I, V$-$I CMD are shown
in Figure \ref{fig_1} assuming $(m-M)_V$ = 14.43 and a reddening
of E(B$-$V) = 0.02 as in BC01 and BCS02.

From each model and a given set of C/N/O abundances, synthetic spectra
were computed using the SSG program
\citep[][and references therein]{1994MNRAS.268..771B}
and the line list of \citet{1995AJ....110.3035T} at a step size
of 0.02\AA\ (see BC01 for further details) assuming the average heavy element
compositions of \citet{1993AJ....106.1490K,1997AJ....113..279K}.
The result was then convolved with a Gaussian kernel to match
the resolution of the observed spectra and I(CH) and $s_{NH}$ indices
were measured.
The model indices for I(CH) are illustrated in Figure \ref{fig_2} for
four C abundances (as in BCS02): \cabund\ = $-$0.85 and \cabund\ = $-$1.1,
which roughly match the observed compositions of M13's CN-weak and
strong bright giants respectively \citep[see][]{1996AJ....112.1511S}, and
\cabund\ = 0.0 and \cabund\ = $-$0.5.
Also plotted in Figure \ref{fig_2} are $s_{NH}$ indices for a variety of
\nabund\ values.
Note that among these relatively warm MS/MSTO stars, there is
little sensitivity in the CH (NH) band strengths to changes in
N, O (C, O) abundances (as opposed to the cool giants
where molecular equilibrium must be considered, particularly with
regard to O).
As a check of this, Table \ref{tbl-1} shows the sensitivity of I(CH) and
$s_{NH}$ to such changes in a cool MS model (T$_{eff}$=5601, log g=4.66,
corresponding to an M13 MS star with $I$=19.60).

Following BCS02, we have applied the method of \citet{1990ApJ...359..307B} to
convert the observed indices to corresponding C and N abundances:
the model isoabundance curves were interpolated to the $M_I$ of each
program star, and the observed index converted into the corresponding
abundance based on the synthetic index at that $M_I$.
Resulting C and N abundances are plotted in Figures \ref{fig_4} and
\ref{fig_5}.
Note that the large error bars which accompany the stars of Figure
\ref{fig_4} near I=19 and the stars of low \cabund\ ($\approx -1$) are
due to the overall weakness of the CH bands --- small errors
in I(CH) therefore result in large changes in \cabund.
Likewise, a similar situation exists among the MSTO stars with
measured NH band strengths.

\subsection{THE BRIGHT STAR SAMPLE}

As with the faint stars, the I(CH) index was measured from the spectra of
the six bright M13 giants.
For each star this value was compared to synthetic indices
generated from model atmospheres whose stellar parameters were taken from
the high resolution analyses of \citet{1993AJ....106.1490K,1997AJ....113..279K},
and \citet{1996AJ....112..545P}, including their heavy element and [O/Fe] abundances.
Where available, \nabund\ values from \citet{1996AJ....112.1511S} were
also used.
For two stars (K188 and III-7), N abundances were not available
from the literature, and a value of +1.0 was assumed.
For star III-7, an [O/Fe] of 0.0 was used.
The model parameters and the resulting \cabund\ abundance which
matched the observed I(CH) indices are listed in Table \ref{tbl-2}
along with the C abundances from \citet{1996AJ....112.1511S} and
\citet{1981ApJS...47....1S}.

For the four bright giants in common with \citet{1996AJ....112.1511S}, we
find an average offset of 0.03($\pm$ 0.14) dex in \cabund\ (present $-$ Smith).
We therefore consider our C abundances to be essentially on the
same scale, as might be expected considering the similar analysis
tools used.
The difference between our results and those of \citet{1981ApJS...47....1S} are
somewhat larger: 0.25($\pm$0.23).
However, almost half of this offset is driven by the
result for II-76.
Excluding this star reduces the average difference to 0.14($\pm$0.07).
Note that II-76 has both a high [O/Fe] and a lower \nabund\ abundance
as might be expected from a star with a lesser amount of CN(O)-cycle
material in its atmosphere \citep[it also has the second lowest
Na abundance of the large sample of][]
{1996AJ....112..545P}.
The source of this discrepancy is likely the cooler model used for
II-76 by Suntzeff (T$_{eff}$ = 4220K versus the 4350K used here),
as well as the lower O abundance ([O/Fe] = 0.0) and higher microturbulent
velocity (2.5 \kms).
Repeating our analysis with the values used by Suntzeff reduces our
resulting \cabund\ by 0.32 dex to $-$0.96.
The luminous stars of Suntzeff plotted in Figure \ref{fig_4} have
therefore been shifted by 0.14 in \cabund\ to place them on our
abundance scale.

Given the use of the same modeling codes, line lists, and CH indices
throughout our analysis, we presume the resulting C abundances from both the faint
and bright star samples, as well as those of \citet{1996AJ....112.1511S}
and \citet{1981ApJS...47....1S} (with the appropriate shift), to be on the
same abundance scale.
Any systematic differences due to different telescope/spectrograph
systems will be minimized by the use of the I(CH) index
which uses continuum bands both blueward and redward of
the CH feature to remove slope differences
due to variations in instrumental response.

\section{RESULTS}

There are several points to be made about the present results,
which are given in tabular form in Tables \ref{tbl-A1} and \ref{tbl-A2} of
Appendix A.
First, as can be seen in Figure \ref{fig_2}, significant differences
in \cabund\ exist among stars at least 1.5 mag fainter than the
MSTO in M13.
This corresponds to a mass of approximately 0.66 $M_\odot$
using the isochrone of Figure \ref{fig_1}.
Among these old MS stars, CN(O)-cycle reactions are entirely confined
to the central core \citep[see for example Figures 4 and 5 of][]{2002ApJ...568..979R}
and as has been pointed out by numerous investigators, MS stars such
as these are not thought to possess a mechanism that connects their
surface with regions of energy generation (namely the core).
Indeed, should such mixing take place, the subsequent paths of the stars
in the CMD would be radically altered by the infusion of fresh H
into the core \citep[e.g.][]{1988PASP..100..314V}.
One must conclude the source of the observed differences in
\cabund\ is likely not the stars themselves.
Moreover, the values of \cabund\ among the MS stars are consistent with
those found by BCS02 among the M13 SGB stars (see Figure \ref{fig_4})
and imply little change in composition has occurred from the MS to at
least the base of the SGB.

Figure \ref{fig_4} also includes the \cabund\ values of \citet[][]{1996AJ....112.1511S}
and \citet[][]{1981ApJS...47....1S} (shifted upwards by 0.14 dex).
As discussed in BCS02, there appears to be a marked decline in
\cabund\ towards higher luminosities among the M13 giants.
Clearly an evolutionary change such as this can be best interpreted
as the result of a mixing process bringing up C-depleted material
from a region in which at least CN-cycle reactions are operating
(see BCS02 for a more detailed discussion).
Also shown in Figure \ref{fig_4} is the location of the LF bump
in M13 \citep[from][]{1998MNRAS.293..434P} --- the event which marks the
destruction of the molecular weight gradient thought to inhibit
deep mixing.
Unfortunately, the luminosity at which the onset of C depletion begins
is uncertain due to the gap in the available data (from 15 $<$ V $<$ 17).
However, since an extrapolation of the trend in
giant-branch [C/Fe] abundance faintward intersects the magnitude of the LF
bump at the average abundance of the fainter stars, it is reasonable to
infer that the abundance decline begins near that event; neither a
significant decrease nor a significant increase in carbon abundance with a
subsequent recovery to the original value hidden within the gap in our data
is reasonably to be expected.

The \nabund\ values determined from the NH band strengths of
the MSTO stars are plotted in Figure \ref{fig_5}.
Although the error bars are admittedly larger than one would like
owing to the weaknesses of the CH and NH bands among the warmer MSTO
stars, a general anticorrelation between \cabund\ and \nabund\ is suggested.
Note that these abundances do not suffer from the inherent
tendency towards C/N anticorrelations of analyses based on
CH and CN band strengths.
Of course an overall C/N anticorrelation is known to
be present among the evolved M13 stars and the values
for the bright RGB stars of \citet{1996AJ....112.1511S} are also shown in
Figure \ref{fig_5}.
Immediately apparent is the shift of the RGB stars towards
lower \cabund, as is expected from Figure \ref{fig_4}.
If C-poor/N-rich material is indeed being circulated into the
stellar envelopes during RGB ascent, the lack of near solar
\nabund\ RGB stars is also explained (although the error bars
on the two lower \nabund\ MSTO stars severely limit the
weight which can be placed on this statement).
That higher N abundances do not appear to be found among
the RGB stars under these circumstances is perhaps not
a surprise if these stars are already leaving the MSTO with large
\nabund\ overabundances: an M13 MSTO star with \cabund\ = $-$0.4
and \nabund\ = 1.0 which undergoes a mixing episode reducing
\cabund\ to $-$1.2 will experience a rise in \nabund\ of only
0.05 dex --- in essence, the N
abundances are already so large, the addition of freshly minted
N via the CN-cycle results in only a small fractional change in
\nabund.
Thus, while the error bars in Figure \ref{fig_5} are large, we
can at least claim it is not inconsistent with the assertion that
we are seeing substantial star-to-star variations in C (and N)
set early in the cluster history, which are further being modified
by mixing during RGB ascent.
The possibility of also mixing ON-cycle material to the surface is more
difficult to assess because of the large N variations among the MSTO
stars.
In the example above, an additional reduction in [O/Fe] from +0.45 to
$-0.35$ would increase \nabund\ by 0.46 dex --- and among the bright
giants, even larger O depletions (as much as [O/Fe] = $-0.7$ to $-0.8$)
have been noted.
Starting with an even larger N overabundance of +1.4 reduces the change
in [N/Fe] to +0.25.
However, at least from the small sample of Figure \ref{fig_5}, it appears
that none of the bright RGB stars possess larger N abundances than their
MSTO counterparts, which in turn suggests the envelopes of at least the
initially N-rich stars may not be cycled through a region of ON-cycle
reactions while on the RGB.
Clearly knowledge of the O abundances of the M13 MSTO stars would
help settle this question.

A similar result was noted in the more metal-poor clusters M92 and
M15 by \citet{1982ApJS...49..207C} and \citet{1983ApJ...266..144T}
(respectively) --- that
substantial N overabundances are present from the SGB to AGB that are not
necessarily correlated with C abundances.
Indeed, an analogous situation can been seen in the present results and those
of \citet{2002AJ....123.2525C} for M5 (see Figures \ref{fig_5} and \ref{fig_6}):
the ``higher'' \cabund\ MSTO stars (at $\approx -0.4$) span almost a dex in
\nabund.
It is clear that if we are to ascribe the same mechanism to the origin of the
SGB/MSTO inhomogeneities in these clusters, it must be operating at
the MSTO or earlier.

\section{DISCUSSION}

That significant and correlated star-to-star differences in C and N, as well
as O, Na, Al, and Mg have been found among the SGB, MSTO,
and MS stars of several clusters (see references above), implies
the operation of some process external to the present stars,
presumably having taken place early in the cluster history.
The discussion of \citet{1998MNRAS.298..601C} includes a comprehensive
look at various possibilities.
It is worth while however, to revisit a few of the more
critical constraints on any theory of the origin of the abundance
variations.

First, whatever mechanism has altered the light-element compositions
of the cluster stars has left the heavy elements essentially untouched,
at least to the limits of our ability to determine them --- the
analysis of M5 by \citet{2003AJ....125..224R} is an excellent example.
This alone would seem to exclude the possibility of the light-element
variations arising from the merger of two distinct proto-cluster
clouds (as has been pointed out by numerous authors).

Second, these abundance variations appear to be almost ubiquitous among the
population of Galactic globular clusters.
To highlight this, we have plotted in Figure \ref{fig_6} the \cabund\ 
and \nabund\ values for the present sample of M13 MSTO stars,
the 47 Tuc MS stars of \citet{Briley-2004}, the M5 SGB stars of
\citet{2002AJ....123.2525C}, and the MSTO stars of M71
from BC01.
Note that BC01 did not directly extract C and N abundances from
their observed indices --- we have converted them here following the
procedure outlined in \citet{2002AJ....123.2525C} and using the
indices and models presented in BC01; the values are given in
Appendix A, Table \ref{tbl-A3}.

Third, the elements which are observed to vary are associated with proton
capture nucleosynthesis under conditions of CN and ON-cycling.
The source/site must process these CNO-group elements and
return this material to the cluster to be incorporated into the present
population of low mass stars either before, during, or after their
formation.

A popular model which fits these constraints is the incorporation of ejecta
from intermediate mass (3-6$M_\odot$) AGB stars undergoing hot
bottom burning and third dredge-up \citep[see][]{2001ApJ...550L..65V},
although difficulties such as the establishment of an O-Na anti-correlation
remain \citep[see for example][]{2003ApJ...590L..99D}.
However, as is discussed in \citet*{2001AJ....122.2561B} and BCS02,
the quantities of material required to produce
the observed star-to-star differences among the low luminosity stars
(most notably extreme C depletions), which are clearly not diluted as the
convective envelopes deepen during RGB ascent, rules out any sort of simple
accretion model.
Indeed, for the present M13 stars, roughly 70\% of a C-poor MS star's total mass
must be captured ejecta if the accreted matter is completely
free of C (see BCS02).
It is of course unclear how such an enormous amount of material
can be returned to the cluster without appealing to a shallow initial
mass function \citep[see][]{2001AJ....122.2561B}, nor how the present stars
can sweep up the necessary mass of ejecta \citep[although a novel look at
accretion by][suggests significant quantities of AGB ejecta could be captured by stars
in clusters with high central concentrations, it should be noted that M13 is definitely {\it not\/} a cluster with a high central concentration] {2002A&A...383..491T}.
We note in Figure \ref{fig_6} that the depletions in C do appear smaller in
the more metal-rich clusters M71 and 47 Tuc in accord with the
prediction the of AGB ejecta models of \citet{2001ApJ...550L..65V}.
Yet at the same time, if one presumes the highest \cabund\ SGB/MSTO stars in M13
and M5 to represent the original (accretion free) C abundance of the cluster stars,
they are still some 0.4 dex more C-poor than their 47 Tuc/M71 counterparts,
implying either truly primordial (i.e., pre-accretion) differences in at least C
or that nearly all the present stars in M13 and M5 have undergone at least
some accretion of C-poor material.
However, the spread in \nabund\ is essentially identical among all four clusters.
Clearly, knowledge of the patterns of [O/Fe] and [Na/Fe] among the
present stars would help constrain the AGB ejecta theories.

An interesting counterpoint to this model is the scenario suggested by
\citet{1982ApJS...49..207C} and \citet{1983ApJ...266..144T}
to explain similar results among M92 and M15 SGB stars ---
that the stars of these clusters were inhomogeneously ``polluted'' 
by an injection of raw C from intermediate
mass AGB stars which is subsequently converted into N in the present
stars before SGB evolution thereby explaining both the C deficiencies and
large N enhancements as well as star-to-star differences in (C+N).
This has the additional advantage of requiring considerably more modest
composition modifications (a factor of 4 or so in C from star to star), which in
turn lowers the mass of captured ejecta required.
However, to explain the large C depletions already in place by the MSTO,
significant processing of the envelope through a region of CN-cycling
must have taken place while the stars occupied the MS. One
then returns to the difficulty of mixing in such stars 
discussed above.

Another site of the proton-capture reactions has recently been suggested
by \citet{Li-Burstein03}, who note that the high mass (250-300 $M_\odot$)
zero metallicity models of \citet{2001ApJ...550..372F} tend to mix He and
He-burning products into their H-burning shells during the later stages of
He-burning.
This fresh C, N, O is partially processed into N while at the same time, the
stars expand into red supergiants.
If mass loss also occurs at this point, the cluster could be seeded with
freshly produced C/O-poor, N-rich material.
Such a scenario is presented within the context of the cluster formation
history of \citet{1986A&A...168...81C} --- that the GCs formed from primordial
material (zero-metal) that was subsequently enriched by the supernovae  
of massive stars before low mass stars could form.
However, the problem remains that the production/seeding and
mixing of the heavy-elements must be decoupled from that of the light-elements
in order to explain the remarkable homogeneity of Fe, Ti, Ca, etc. within the GCs.
In the context of GC formation in a well mixed supershell
\citep[e.g.][]{1991ApJ...376..115B}
this is difficult to explain if the CNO-modified material is ejected prior to the driving
supernovae and subsequent supershell expansion/mixing.

\acknowledgments
The entire Keck/LRIS user community owes a huge debt
to Jerry Nelson, Gerry Smith, Bev Oke, and many other people who have
worked to make the Keck Telescope and LRIS a reality.  We are grateful
to the W. M. Keck Foundation, and particularly its late president,
Howard Keck, for the vision to fund the construction of the W. M. Keck
Observatory. We also wish to express our thanks to Roger Bell whose
SSG code was instrumental in this project and the anonymous referee
for their suggestions.
Partial support was provided by the National Science Foundation under
grant AST-0098489 to MMB and grant AST-0205951 to JGC and by
the F. John Barlow professorship and UW Oshkosh Faculty
Development Program (MMB).

\appendix
\section{TABLES OF OBSERVED INDICES AND RESULTING
ABUNDANCES}

\begin{deluxetable}{cccccccc}
\tablenum{3}
\tablecaption{Current Program Stars: Photometry, Indices, and Abundances
\label{tbl-A1}}
\tablewidth{0pt}
\tablehead{
\colhead{Star} &
\colhead{V} &
\colhead{I} &
\colhead{V$-$I} &
\colhead{I(CH)} &
\colhead{$s_{NH}$} &
\colhead{[C/Fe]} &
\colhead{[N/Fe]}
}
\startdata
41211\_2349 & 17.30 & 17.10 & 0.20 & -0.010 & -0.024 & - & - \\
41217\_2535 & 17.63 & 16.88 & 0.75 & 0.104 & 0.228 & -0.49 & 1.09 \\
41132\_2535 & 18.20 & 17.61 & 0.59 & 0.013 & 0.060 & - & 1.37 \\
41185\_2646 & 18.52 & 17.98 & 0.54 & 0.008 & 0.031 & - & 1.24 \\
41165\_2813 & 18.64 & 18.10 & 0.54 & 0.034 & -0.026 & -0.25 & -0.14 \\
41204\_2622 & 18.89 & 18.34 & 0.55 & 0.034 & 0.004 & -0.39 & 0.81 \\
41302\_2212 & 18.90 & 18.32 & 0.58 & 0.057 & - & 0.30 & - \\
41228\_2301 & 19.05 & 18.53 & 0.52 & 0.022 & 0.054 & - & 1.20 \\
41211\_2513 & 19.06 & 18.49 & 0.57 & 0.032 & -0.022 & -0.50 & 0.02 \\
41157\_2535 & 19.07 & 18.49 & 0.58 & 0.028 & 0.069 & -0.77 & 1.32 \\
41170\_2333 & 19.07 & 18.50 & 0.57 & 0.024 & 0.094 & - & 1.46 \\
41191\_2527 & 19.38 & 18.77 & 0.61 & 0.030 & - & -0.91 & - \\
41274\_2133 & 19.41 & 18.73 & 0.68 & 0.000 & - & - & - \\
41265\_2254 & 19.55 & 18.94 & 0.61 & 0.061 & - & -0.12 & - \\
41197\_2648 & 19.58 & 18.97 & 0.61 & 0.038 & 0.175 & -0.66 & 1.47 \\
41171\_2802 & 19.92 & 19.33 & 0.59 & 0.043 & - & -0.93 & - \\
41173\_2418 & 19.97 & 19.30 & 0.67 & 0.097 & - & -0.16 & - \\
41188\_2516 & 19.98 & 19.31 & 0.67 & 0.089 & - & -0.24 & - \\
41107\_2643 & 20.00 & 19.33 & 0.67 & 0.071 & - & -0.44 & - \\
41334\_2223 & 20.00 & 19.35 & 0.65 & 0.070 & - & -0.47 & - \\
41243\_2227 & 20.01 & 19.39 & 0.62 & 0.043 & - & -1.02 & - \\
41122\_2707 & 20.03 & 19.36 & 0.67 & 0.100 & - & -0.21 & - \\
41227\_2356 & 20.04 & 19.30 & 0.74 & 0.102 & - & -0.12 & - \\
41120\_2625 & 20.05 & 19.36 & 0.69 & 0.032 & - & -1.47 & - \\
41264\_2258 & 20.05 & 19.31 & 0.74 & 0.070 & - & -0.43 & - \\
41296\_2221 & 20.05 & 19.40 & 0.65 & 0.113 & - & -0.15 & - \\
41135\_2723 & 20.06 & 19.38 & 0.68 & 0.101 & - & -0.22 & - \\
41277\_2355 & 20.11 & 19.35 & 0.76 & 0.049 & - & -0.78 & - \\
41198\_2646 & 20.13 & 19.42 & 0.71 & 0.075 & - & -0.49 & - \\
41219\_2530 & 20.15 & 19.48 & 0.67 & 0.037 & - & -1.41 & - \\
41167\_2650 & 20.19 & 19.50 & 0.69 & 0.115 & - & -0.25 & - \\
41276\_2132 & 20.21 & 19.60 & 0.61 & 0.077 & - & -0.60 & - \\
41202\_2321 & 20.23 & 19.55 & 0.68 & 0.056 & - & -0.87 & - \\
41210\_2353 & 20.23 & 19.56 & 0.67 & 0.127 & - & -0.23 & - \\
41130\_2540 & 20.24 & 19.54 & 0.70 & 0.077 & - & -0.56 & - \\
41340\_2151 & 20.24 & 19.55 & 0.69 & 0.104 & - & -0.38 & - \\
41130\_2641 & 20.25 & 19.54 & 0.71 & 0.130 & - & -0.19 & - \\
41131\_2639 & 20.26 & 19.50 & 0.76 & 0.086 & - & -0.47 & - \\
41207\_2619 & 20.26 & 19.56 & 0.70 & 0.113 & - & -0.33 & - \\
41243\_2229 & 20.27 & 19.56 & 0.71 & 0.096 & - & -0.45 & - \\
41279\_2407 & 20.27 & 19.60 & 0.67 & 0.095 & - & -0.49 & - \\
41191\_2530 & 20.28 & 19.54 & 0.74 & 0.113 & - & -0.31 & - \\
41208\_2557 & 20.31 & 19.60 & 0.71 & 0.087 & - & -0.53 & - \\
41255\_2229 & 20.32 & 19.55 & 0.77 & 0.103 & - & -0.39 & - \\
41278\_2232 & 20.32 & 19.64 & 0.68 & 0.097 & - & -0.50 & - \\
41301\_2213 & 20.32 & 19.65 & 0.67 & 0.125 & - & -0.35 & - \\
41315\_2143 & 20.33 & 19.62 & 0.71 & 0.135 & - & -0.25 & - \\
41270\_2213 & 20.35 & 19.60 & 0.75 & 0.105 & - & -0.43 & - \\
41166\_2612 & 20.38 & 19.65 & 0.73 & 0.141 & - & -0.24 & - \\
41183\_2653 & 20.40 & 19.66 & 0.74 & 0.118 & - & -0.41 & - \\
41255\_2307 & 20.46 & 19.66 & 0.80 & 0.053 & - & -1.15 & -
\enddata
\end{deluxetable}

\begin{deluxetable}{cccccc}
\tablenum{4}
\tablecaption{M5 Subgiants from BCS02: Program Stars, Photometry, Indices, and
Abundances
\label{tbl-A2}}
\tablewidth{0pt}
\tablehead{
\colhead{Star} &
\colhead{V} &
\colhead{I} &
\colhead{V$-$I} &
\colhead{I(CH)} &
\colhead{[C/Fe]}
}
\startdata
41230\_2604 & 16.83 & 16.00 & 0.83 & 0.166 & -0.34 \\
41244\_2423 & 16.88 & 16.05 & 0.83 & 0.182 & -0.25 \\
41224\_2734 & 16.92 & 16.11 & 0.81 & 0.154 & -0.40 \\
41299\_2630 & 16.95 & 16.12 & 0.83 & 0.212 & -0.07 \\
41213\_2642 & 16.99 & 16.20 & 0.79 & 0.098 & -0.72 \\
41259\_2821 & 17.03 & 16.23 & 0.80 & 0.182 & -0.22 \\
41210\_2830 & 17.07 & 16.30 & 0.77 & 0.168 & -0.29 \\
41320\_2941 & 17.07 & 16.31 & 0.76 & 0.118 & -0.56 \\
41207\_2719 & 17.09 & 16.31 & 0.78 & 0.136 & -0.47 \\
41296\_2957 & 17.11 & 16.34 & 0.77 & 0.201 & -0.10 \\
41249\_2549 & 17.30 & 16.46 & 0.84 & 0.104 & -0.63 \\
41210\_2834 & 17.05 & 16.48 & 0.57 & 0.128 & -0.48 \\
41301\_2440 & 17.32 & 16.49 & 0.83 & 0.208 & -0.03 \\
41188\_2619 & 17.34 & 16.57 & 0.77 & 0.081 & -0.84 \\
41260\_3026 & 17.35 & 16.60 & 0.75 & 0.112 & -0.55 \\
41212\_2744 & 17.39 & 16.60 & 0.79 & 0.165 & -0.24 \\
41256\_2801 & 17.43 & 16.66 & 0.77 & 0.082 & -0.79 \\
41260\_2850 & 17.43 & 16.69 & 0.74 & 0.127 & -0.43 \\
41340\_2401 & 17.54 & 16.72 & 0.82 & 0.072 & -0.92 \\
41252\_2524 & 17.53 & 16.73 & 0.80 & 0.149 & -0.28 \\
41282\_2908 & 17.49 & 16.76 & 0.73 & 0.094 & -0.62 \\
41284\_2922 & 17.54 & 16.81 & 0.73 & 0.113 & -0.47 \\
41217\_2534 & 17.63 & 16.88 & 0.75 & 0.106 & -0.47 \\
41260\_2459 & 17.69 & 16.88 & 0.81 & 0.116 & -0.40 \\
41256\_3005 & 17.67 & 16.94 & 0.73 & 0.023 & -\tablenotemark{a} \\
41249\_2548 & 18.09 & 17.03 & 1.06 & -0.032 & -\tablenotemark{a} \\
42103\_2722 & - & 17.03 & - & 0.091 & -\tablenotemark{a} \\
41253\_2521 & 17.79 & 17.07 & 0.72 & 0.068 & -\tablenotemark{a} \\
42104\_2748 & - & 17.15 & - & 0.067 & -\tablenotemark{a} \\
42108\_2808 & - & 17.18 & - & 0.039 & -\tablenotemark{a} \\
42071\_2457 & - & 17.20 & - & 0.051 & -\tablenotemark{a} \\
42055\_2321 & - & 17.23 & - & 0.053 & -\tablenotemark{a} \\
42097\_2610 & - & 17.24 & - & 0.042 & -\tablenotemark{a} \\
42088\_2635 & - & 17.33 & - & 0.046 & -\tablenotemark{a} \\
42077\_2623 & - & 17.40 & - & 0.033 & -\tablenotemark{a} \\
41112\_2621 & 18.05 & 17.44 & 0.61 & 0.012 & -\tablenotemark{a} \\
41284\_2930 & 18.02 & 17.47 & 0.55 & 0.031 & -\tablenotemark{a} \\
42062\_2223 & - & 17.51 & - & 0.030 & -\tablenotemark{a} \\
42095\_2656 & - & 17.52 & - & 0.101 & -\tablenotemark{a} \\
42104\_2854 & - & 17.55 & - & 0.009 & -\tablenotemark{a} \\
42068\_2553 & - & 17.60 & - & 0.017 & -\tablenotemark{a} \\
42040\_2211 & - & 17.63 & - & 0.048 & -\tablenotemark{a} \\
42073\_2606 & - & 17.68 & - & 0.016 & -\tablenotemark{a} \\
42028\_2442 & - & 17.71 & - & 0.006 & -\tablenotemark{a} \\
42033\_2200 & - & 17.72 & - & 0.015 & -\tablenotemark{a} \\
42072\_2737 & 18.32 & 17.74 & 0.58 & 0.035 & -\tablenotemark{a} \\
42071\_2508 & - & 17.74 & - & 0.010 & -\tablenotemark{a} \\
42068\_2541 & - & 17.76 & - & 0.011 & -\tablenotemark{a} \\
42129\_2833 & - & 17.78 & - & 0.039 & -\tablenotemark{a} \\
42096\_2708 & - & 17.79 & - & 0.047 & -\tablenotemark{a} \\
42071\_2515 & - & 17.92 & - & -0.002 & -\tablenotemark{a} \\
41281\_2911 & 18.15 & 17.96 & 0.19 & 0.001 & -\tablenotemark{a} \\
42051\_2422 & - & 17.96 & - & 0.018 & -\tablenotemark{a} \\
42062\_2405 & - & 17.96 & - & 0.044 & -\tablenotemark{a} \\
42064\_2355 & - & 17.96 & - & 0.044 & -\tablenotemark{a} \\
42078\_2253 & - & 17.99 & - & 0.014 & -\tablenotemark{a} \\
41213\_2651 & 18.54 & 18.00 & 0.54 & 0.020 & -\tablenotemark{a} \\
42060\_2650 & - & 18.04 & - & 0.013 & -\tablenotemark{a} \\
42034\_2250 & - & 18.04 & - & 0.043 & -\tablenotemark{a} \\
42065\_2305 & - & 18.05 & - & 0.029 & -\tablenotemark{a} \\
42029\_2445 & - & 18.07 & - & 0.013 & -\tablenotemark{a} \\
42050\_2430 & - & 18.10 & - & 0.028 & -\tablenotemark{a} \\
42058\_2452 & - & 18.17 & - & -0.002 & -\tablenotemark{a} \\
41099\_2615 & 18.75 & 18.19 & 0.56 & -0.006 & -\tablenotemark{a} \\
42064\_2349 & - & 18.21 & - & -0.009 & -\tablenotemark{a} \\
41341\_2405 & 18.83 & 18.22 & 0.61 & 0.004 & -\tablenotemark{a} \\
42035\_2345 & - & 18.24 & - & 0.012 & -\tablenotemark{a} \\
42099\_2820 & - & 18.36 & - & 0.035 & -\tablenotemark{a} \\
41136\_2455 & 19.01 & 18.42 & 0.59 & 0.028 & -\tablenotemark{a} \\
41081\_2630 & 19.05 & 18.46 & 0.59 & 0.039 & -\tablenotemark{a} \\
41157\_2535 & 19.07 & 18.49 & 0.58 & 0.024 & -\tablenotemark{a} \\
41228\_2301 & 19.05 & 18.53 & 0.52 & 0.020 & -\tablenotemark{a} \\
41121\_2548 & 19.12 & 18.54 & 0.58 & 0.009 & -\tablenotemark{a} \\
41102\_2643 & 19.14 & 18.54 & 0.60 & 0.008 & -\tablenotemark{a} \\
41096\_2617 & 19.14 & 18.56 & 0.58 & 0.130 & -\tablenotemark{a} \\
41173\_2525 & 19.15 & 18.56 & 0.59 & 0.040 & -\tablenotemark{a} \\
41150\_2415 & 19.16 & 18.56 & 0.60 & 0.021 & -\tablenotemark{a} \\
41172\_2423 & 19.17 & 18.58 & 0.59 & 0.046 & -\tablenotemark{a} \\
41216\_2412 & 19.18 & 18.61 & 0.57 & 0.034 & -\tablenotemark{a} \\
41116\_2614 & 19.21 & 18.62 & 0.59 & 0.023 & -\tablenotemark{a} \\
41280\_2145 & 19.21 & 18.62 & 0.59 & 0.034 & -\tablenotemark{a} \\
41139\_2533 & 19.21 & 18.62 & 0.59 & 0.041 & -\tablenotemark{a} \\
41317\_2148 & 19.21 & 18.66 & 0.55 & 0.038 & -\tablenotemark{a} \\
41256\_2223 & 19.22 & 18.66 & 0.56 & 0.040 & -\tablenotemark{a} \\
41204\_2446 & 19.27 & 18.66 & 0.61 & 0.054 & -\tablenotemark{a} \\
41279\_2334 & 19.27 & 18.67 & 0.60 & 0.044 & -\tablenotemark{a} \\
41200\_2315 & 19.28 & 18.68 & 0.60 & 0.060 & -\tablenotemark{a} \\
41263\_2211 & 19.28 & 18.69 & 0.59 & 0.021 & -\tablenotemark{a} \\
41236\_2215 & 19.37 & 18.77 & 0.60 & 0.019 & -\tablenotemark{a} \\
41255\_2324 & 19.39 & 18.79 & 0.60 & 0.023 & -\tablenotemark{a} \\
41174\_2358 & 19.39 & 18.79 & 0.60 & 0.030 & -\tablenotemark{a} \\
41134\_2417 & 19.41 & 18.80 & 0.61 & 0.017 & -\tablenotemark{a} \\
41231\_2331 & 19.42 & 18.81 & 0.61 & 0.048 & -\tablenotemark{a} \\
41094\_2540 & 19.43 & 18.82 & 0.61 & 0.050 & -\tablenotemark{a} \\
41136\_2550 & 19.45 & 18.85 & 0.60 & -0.005 & -\tablenotemark{a} \\
41310\_2206 & 19.65 & 18.88 & - & 0.030 & -\tablenotemark{a} \\
41290\_2218 & 19.49 & 18.89 & 0.60 & 0.016 & -\tablenotemark{a} \\
41219\_2401 & 19.50 & 18.96 & 0.54 & 0.010 & -\tablenotemark{a} \\
41161\_2532 & 19.80 & 19.18 & 0.62 & 0.029 & -\tablenotemark{a} \\
41175\_2421 & 19.93 & 19.25 & 0.68 & 0.085 & -\tablenotemark{a} \\
41286\_2327 & 20.13 & 19.40 & 0.73 & 0.090 & -\tablenotemark{a} \\
41140\_2549 & 20.23 & 19.52 & 0.71 & 0.093 & -\tablenotemark{a} \\
41266\_2209 & 20.43 & 19.71 & 0.72 & 0.093 & -\tablenotemark{a}
\enddata
\tablenotetext{a}{Values of \cabund ~ were not determined for stars near the MSTO
due to the weakness of the CH-bands.}
\end{deluxetable}

\begin{deluxetable}{ccccccc}
\tablenum{5}
\tablecaption{M71 Subgiants from BC01: Program Stars, Photometry, Indices, and Abundances
\label{tbl-A3}}
\tablewidth{0pt}
\tablehead{
\colhead{Star} &
\colhead{R} &
\colhead{B$-$R} &
\colhead{I(CH)} &
\colhead{S(3839)} &
\colhead{[C/Fe]} &
\colhead{[N/Fe]} 
}
\startdata
C51228\_3737  & 17.00 & 1.38 & 0.138 & 0.131 & -0.17 & 0.32 \\
C51265\_3739  & 17.01 & 1.31 & 0.122 & 0.141 & -0.25 & 0.49 \\
C51314\_3755  & 17.01 & 1.34 & 0.096 & 0.340 & -0.39 & 1.41 \\
C51385\_4166  & 17.01 & 1.40 & 0.092 & 0.378 & -0.40 & 1.54 \\
C51312\_3634  & 17.03 & 1.35 & 0.156 & 0.388 & 0.04 & 1.19 \\
C51418\_4158  & 17.03 & 1.36 & 0.091 & 0.347 & -0.39 & 1.48 \\
C51419\_3870  & 17.03 & 1.38 & 0.112 & 0.333 & -0.24 & 1.29 \\
C51291\_3655  & 17.05 & 1.33 & 0.125 & 0.296 & -0.13 & 1.11 \\
C51417\_3943  & 17.05 & 1.41 & 0.146 & 0.124 & -0.03 & 0.25 \\
C51396\_4020  & 17.11 & 1.33 & 0.140 & 0.327 & 0.04 & 1.15 \\
C51285\_3749  & 17.12 & 1.29 & 0.120 & 0.142 & -0.10 & 0.55 \\
C51260\_4161  & 17.13 & 1.40 & 0.096 & 0.351 & -0.20 & 1.49 \\
C51266\_3848  & 17.13 & 1.36 & 0.090 & 0.280 & -0.28 & 1.34 \\
C51254\_3957  & 17.14 & 1.42 & 0.143 & 0.143 & 0.06 & 0.43 \\
C51413\_4033  & 17.14 & 1.27 & 0.092 & 0.236 & -0.27 & 1.18 \\
C51352\_4055  & 17.15 & 1.31 & 0.127 & 0.115 & -0.02 & 0.30 \\
C51424\_3823  & 17.15 & 1.31 & 0.086 & 0.279 & -0.29 & 1.38 \\
C51400\_3529  & 17.16 & 1.31 & 0.118 & 0.105 & -0.08 & 0.26 \\
C51250\_3763  & 17.17 & 1.39 & 0.136 & 0.107 & 0.05 & 0.18 \\
C51373\_3631  & 17.17 & 1.31 & 0.128 & 0.102 & -0.01 & 0.17 \\
C51368\_4074  & 17.18 & 1.31 & 0.077 & 0.253 & -0.37 & 1.39 \\
C51306\_3738  & 17.19 & 1.32 & 0.124 & 0.113 & -0.01 & 0.31 \\
C51277\_3950  & 17.20 & 1.34 & 0.094 & 0.255 & -0.20 & 1.25 \\
C51270\_3931  & 17.21 & 1.40 & 0.082 & 0.285 & -0.27 & 1.43 \\
C51386\_3659  & 17.21 & 1.28 & 0.123 & 0.103 & -0.01 & 0.21 \\
C51416\_3834  & 17.22 & 1.31 & 0.118 & 0.124 & -0.03 & 0.45 \\
C51266\_4149  & 17.24 & 1.34 & 0.092 & 0.273 & -0.18 & 1.32 \\
C51346\_4124  & 17.24 & 1.39 & 0.133 & 0.340 & 0.10 & 1.26 \\
C51267\_4025  & 17.25 & 1.35 & 0.127 & 0.121 & 0.04 & 0.38 \\
C51378\_3975  & 17.25 & 1.34 & 0.124 & 0.106 & 0.02 & 0.25 \\
C51404\_3918  & 17.25 & 1.35 & 0.111 & 0.109 & -0.07 & 0.35 \\
C51308\_3765  & 17.26 & 1.27 & 0.099 & 0.283 & -0.12 & 1.31 \\
C51316\_3960  & 17.26 & 1.29 & 0.071 & 0.276 & -0.37 & 1.53 \\
C51377\_3737  & 17.26 & 1.25 & 0.105 & 0.122 & -0.10 & 0.51 \\
C51430\_3648  & 17.29 & 1.34 & 0.083 & 0.234 & -0.25 & 1.27 \\
C51385\_3962  & 17.30 & 1.24 & 0.107 & 0.113 &  -  &  -  \\
C51290\_3644  & 17.33 & 1.36 & 0.113 & 0.109 & -0.03 & 0.36 \\
C51224\_4027  & 17.35 & 1.31 & 0.098 & 0.161 & -0.12 & 0.84 \\
C51279\_3957  & 17.35 & 1.36 & 0.120 & 0.139 & 0.02 & 0.57 \\
C51287\_4050  & 17.35 & 1.38 & 0.116 & 0.135 & 0.00 & 0.55 \\
C51261\_4172  & 17.36 & 1.32 & 0.121 & 0.108 & 0.03 & 0.30 \\
C51281\_3638  & 17.37 & 1.28 & 0.111 & 0.113 & -0.04 & 0.41 \\
C51331\_4042  & 17.38 & 1.28 & 0.076 & 0.229 & -0.30 & 1.34 \\
C51252\_4126  & 17.39 & 1.39 & 0.098 & 0.134 & -0.12 & 0.68 \\
C51279\_3842  & 17.39 & 1.28 & 0.124 & 0.082 & 0.05 & -0.14 \\
C51310\_3849  & 17.39 & 1.29 & 0.109 & 0.140 & -0.04 & 0.64 \\
C51334\_3732  & 17.40 & 1.29 & 0.115 & 0.112 & 0.00 & 0.37 \\
C51262\_3874  & 17.42 & 1.33 & 0.114 & 0.103 & -0.01 & 0.29 \\
C51377\_3766  & 17.42 & 1.29 & 0.080 & 0.231 & -0.25 & 1.31 \\
C51243\_4217  & 17.43 & 1.35 & 0.072 & 0.211 & -0.34 & 1.31 \\
C51412\_4143  & 17.43 & 1.27 & 0.089 & 0.170 &  -  &  - \\
C51287\_3658  & 17.44 & 1.30 & 0.114 & 0.099 & -0.01 & 0.23 \\
C51315\_4161  & 17.44 & 1.30 & 0.134 & 0.458 & 0.22 & 1.60 \\
C51324\_3542  & 17.45 & 1.38 & 0.068 & 0.236 & -0.38 & 1.45 \\
C51405\_3749  & 17.45 & 1.24 & 0.072 & 0.234 & -0.31 & 1.39 \\
C51252\_3923  & 17.46 & 1.29 & 0.081 & 0.253 & -0.23 & 1.38 \\
C51235\_3931  & 17.47 & 1.34 & 0.106 & 0.106 & -0.06 & 0.37 \\
C51244\_3757  & 17.47 & 1.35 & 0.116 & 0.126 & 0.01 & 0.50 \\
C51296\_3969  & 17.48 & 1.25 & 0.113 & 0.086 & -0.02 & 0.02 \\
C51279\_4119  & 17.49 & 1.36 & 0.123 & 0.105 & 0.06 & 0.25 \\
C51289\_3768  & 17.49 & 1.23 & 0.139 & 0.261 & 0.18 & 1.01 \\
C51416\_4137  & 17.51 & 1.39 & 0.061 & 0.207 & -0.48 & 1.44 \\
C51229\_3628  & 17.52 & 1.28 & 0.082 & 0.240 & -0.23 & 1.33 \\
C51391\_3723  & 17.52 & 1.26 & 0.132 & 0.163 & 0.12 & 0.65 \\
C51393\_4120  & 17.52 & 1.27 & 0.115 & 0.135 & 0.00 & 0.58 \\
C51288\_4040  & 17.53 & 1.30 & 0.130 & 0.140 & 0.11 & 0.52 \\
C51292\_3964  & 17.54 & 1.42 & 0.107 & 0.117 & -0.05 & 0.48 \\
C51398\_3758  & 17.54 & 1.24 & 0.078 & 0.226 & -0.27 & 1.31 \\
C51336\_3569  & 17.55 & 1.23 & 0.107 & 0.110 & -0.06 & 0.41 \\
C51338\_3624  & 17.55 & 1.28 & 0.107 & 0.137 & -0.05 & 0.64 \\
C51402\_3627  & 17.55 & 1.30 & 0.118 & 0.101 & 0.02 & 0.23 \\
C51409\_4045  & 17.56 & 1.28 & 0.108 & 0.105 & -0.05 & 0.35 \\
C51275\_3873  & 17.57 & 1.32 & 0.073 & 0.227 & -0.33 & 1.37 \\
C51299\_4161  & 17.57 & 1.28 & 0.118 & 0.084 & 0.02 & -0.05 \\
C51307\_3821  & 17.59 & 1.28 & 0.082 & 0.232 & -0.24 & 1.30 \\
C51419\_3843  & 17.59 & 1.29 & 0.074 & 0.239 & -0.32 & 1.41
\enddata
\end{deluxetable}

\clearpage
\section{UPDATE ON ANOMALOUS STARS PREVIOUSLY OBSERVED IN M5}

In \citet{2002AJ....123.2525C}, we studied the CH bands in 
a large sample of stars in M5.  Even taking into account the substantial
star-to-star variation seen among the CH band strengths
of the stars in our sample, we denoted six of these stars
as anomalous.  Since that time, we have checked the data
for these stars yet again.  We have found that two of the
six stars were misidentified. C18206\_0533 with  V=18.42
is actually C18188\_0733, with BVI = 17.71, 17.03, 16.17.  
With this correction, as compared to the bulk of our M5 sample
\citep[see Figures~8 and 9 of][]{2002AJ....123.2525C}
the star has normal CH for its
\teff, although its uvCN is still anomalously strong,
but not as much as previously. 
Also, star C18211\_0559  (V=18.06) is actually
C18191\_0559, with BVI = 18.27, 17.57, 16.74.  
Its CN is now reasonable for its corrected $V$ mag, but its
CH index is still unexpectedly strong.

In our earlier paper, we presented low
accuracy radial velocities from the LRIS spectra at H$\alpha$
which 
suggested that 4 of the 6 stars
classified as anomalous are radial velocity members
of M5.  Such data was not available for one star, while the
radial velocity of C18211\_0559
(now identified as C18191\_0559) was 25 km/sec higher than that 
of the cluster.

To verify the membership of C18191\_0559 in M5, we obtained 
low SNR spectra with HIRES \citep{1994SPIE.2198..362V} for it
and for a second star from the LRIS sample.
A single 1200 sec exposure for each was made on
May 1, 2002, a night with considerable clouds.  The HIRES
slit for one of these two also included a second M5 star.
The heliocentric
radial velocities for these three stars derived from the NaD lines
are presented in Table \ref{tbl-B1}.
The radial velocity for M5 found by \citet{2002AJ....123.3277R}
from an extensive high dispersion analysis of stars over
a wide range in luminosity is +55.0 km/sec, so we conclude all three of these stars
are members.

\begin{deluxetable}{cccc}
\tablenum{6}
\tablecaption{Precision Radial Velocities for Three M5 Stars
\label{tbl-B1}}
\tablewidth{0pt}
\tablehead{
\colhead{Star} &
\colhead{V}                  &
\colhead{Radial Velocity}                  &
\colhead{Comment} \\
\colhead{} &
\colhead{mag}                  &
\colhead{km s$^{-1}$}                  &
\colhead{}
}
\startdata
C18225\_0537 & 17.07 & +58.2 & Anomalous star in Cohen et al. \\
C18191\_0554 & 17.12 & +63.1 & in LRIS sample, but not anomalous \\
C18191\_0558 & 17.57 & +58.5 & Anomalous in Cohen et al. as C18211\_0559
\enddata
\end{deluxetable}

\clearpage
%figcaptions

\begin{figure}
\epsscale{0.7}
% \plotone{Briley.fig1.eps}
\caption[Briley.fig1.eps]{The I, V$-$I color-magnitude diagram of M13 is plotted 
using the database of Stetson with the
locations of our program stars and those of BCS02 indicated. Also shown are the positions
of the model points used in the present analysis.
\label{fig_1}}
\end{figure}

\begin{figure}
\epsscale{0.7}
% \plotone{Briley.fig2.eps}
\caption[Briley.fig2.eps]{G-band indices (I(CH)) are plotted as a function of
I for the program stars as well as those of BCS02 (lower panel). Measured
values of $s_{NH}$ are shown in the upper panel. Error bars are one sigma
levels determined from Poisson statistics. Also plotted are model indices
for a variety of \cabund\ and \nabund\ values as discussed in the text.
\label{fig_2}}
\end{figure}

\begin{figure}
\epsscale{0.7}
% \plotone{Briley.fig3.eps}
\caption[Briley.fig3.3ps]{Sample spectra of the NH region of two similar M13 MSTO stars
(left) and the G-band (CH) region of two MS stars. In both, significant
differences are apparent.
\label{fig_3}}
\end{figure}

\begin{figure}
\epsscale{0.7}
% \plotone{Briley.fig4.eps}
\caption[Briley.fig4.eps]{Derived C abundances for the present M13 MS and RGB
stars as well as the SGB stars of BCS02. Also plotted are the \cabund\ 
values from \citet{1996AJ....112.1511S} and \citet{1981ApJS...47....1S}
(the later having been shifted upwards by 0.14 dex as discussed in the
text, the size of this shift is indicated by the lines attached to the symbols). The dashed line indicates the location of the LF bump from 
\citet{1998MNRAS.293..434P} - the point before which mixing is believed
to be inhibited. There is a clear and significant scatter in C abundances
among both the present MS sample and the SGB stars of BCS02.
\cabund\ appears to decrease with V among
the most luminous giants as would be expected from mixing, but the onset
is uncertain.
\label{fig_4}}
\end{figure}

\begin{figure}
\epsscale{0.7}
% \plotone{Briley.fig5.eps}
\caption[Briley.fig5.eps]{Values of \nabund\ are plotted versus \cabund\ for
the M13 MSTO stars where (despite the large error bars) an anticorrelation is suggested.
Also shown are the abundances from luminous giants
from \citet{1996AJ....112.1511S} which, as expected from Figure \ref{fig_4},
appear more deficient in \cabund.
That the presumably mixed RGB stars do not show greater N abundances
than their MSTO counterparts appears to be the result of the large initial
N abundances as discussed in the text.
\label{fig_5}}
\end{figure}

\begin{figure}
\epsscale{0.7}
% \plotone{Briley.fig6.eps}
\caption[Briley.fig6.eps]{The \nabund\ versus \cabund\ values are plotted for MS, MSTO,
and SGB stars in four different clusters.
The present MSTO abundances appear consistent with the SGB stars of
M5 \citep{2002AJ....123.2525C}, a cluster of roughly similar metallicity ([Fe/H] = $-$1.26
versus $-$1.51), as opposed to those of the higher metallicity 47 Tuc and M71
stars ([Fe/H] = $-$0.7).
\label{fig_6}}
\end{figure}

\clearpage

% Main Tables

\begin{deluxetable}{ccccc}
\tablenum{1}
\tablecaption{Indices for Deviations from Assumed Composition for
T$_{eff}$=5601K, log g=4.66 Model
\label{tbl-1}}
\tablewidth{0pt}
\tablehead{
\colhead{[C/Fe]} &
\colhead{[N/Fe]}                  &
\colhead{[O/Fe]}                  &
\colhead{I(CH)}          & 
\colhead{$s_{NH}$}
}

\startdata
-0.50  & 0.0 & +0.40 & 0.075 & 0.044 \\
-0.50  & 1.0 & +0.40 & 0.074 & 0.232 \\
-0.50  & 0.0 & 0.00 & 0.079 & 0.048 \\
-1.00  & 0.0 & +0.40 & 0.042 & 0.044
\enddata
\end{deluxetable}

\begin{deluxetable}{lcccccccccc}
\tablenum{2}
\tablecaption{Indices, Model Atmosphere Parameters, and
Resulting [C/Fe] Abundances for M13 Bright Giants
\label{tbl-2}}
\tablewidth{0pt}
\tablehead{
\colhead{Star} &
\colhead{I(CH)} &
\colhead{T$_{eff}$} &
\colhead{log g} & 
\colhead{$v_t$} &
\colhead {V} &
\colhead {[O/Fe]} &
\colhead {[N/Fe]} &
\colhead {[C/Fe]} &
\colhead {[C/Fe]} &
\colhead {[C/Fe]}
\\
\colhead{} &
\colhead{} &
\colhead{(K)} &
\colhead{} & 
\colhead{km s$^{-1}$} &
\colhead {mag} &
\colhead {} &
\colhead {} &
\colhead {Present} &
\colhead {\citet{1996AJ....112.1511S}} &
\colhead {\citet{1981ApJS...47....1S}}
}

\startdata
IV-25/L-954 & 0.166 & 4000 & 0.15 & 2.25 & 12.09 & $-$0.90 & +1.22 & $-$1.31 & $-$1.36 & - \\
II-67/L-70 & 0.165 & 3950 & 0.20 & 2.10 & 12.12 & $-$0.79 & +1.33 & $-$1.32 & $-$1.34 & - \\
II-76/L-96 & 0.200 & 4350 & 1.15 & 1.85 & 12.52 & +0.46 & +0.59 & $-$0.62 & $-$0.82 & $-$1.2 \\
III-18/L-77 & 0.156 & 4350 & 1.15 & 1.85 & 12.77 & $-$0.18 & +1.10 & $-$1.11 & $-$0.97 & $-$1.3 \\
K188/A1 & 0.238 & 4550 & 1.50 & 1.80 & 13.39 & +0.45 & +1.00 & $-$0.32 & - & $-$0.5 \\
III-7/L-114 & 0.173 & 4600 & 1.65 & 2.00 & 13.45 & 0.00 & +1.00 & $-$0.83 & - & -0.9 \\
\enddata
\end{deluxetable}

\end{document}